# Malak: AI-based multilingual personal assistant to combat misinformation and generative AI safety issues

Farnaz Farid and Farhad Ahamed
Western Sydney University
Farnaz.farid@westernsydney.edu.au, farhad.ahamed@westernsydney.edu.au

**Abstract:** The widespread use of AI technologies to generate digital content has led to increased misinformation and online harm. Deep fake technologies, a type of AI, make it easier to create convincing but fake content on social media, leading to various cyber threats. Malicious actors exploit AI capabilities, posing digital, physical, and psychological harm to individuals. While social media platforms have safety measures such as content rating and feedback systems, these are often used by people with higher digital literacy. There is a lack of preventive measures and a need for user-friendly tools that can be used by people with lower digital literacy. Our goal is to create a user-friendly multilingual AI-based personal assistant, Malak, to reduce online harm and promote safe online interactions, benefiting users with lower literacy levels.

**Research goals and problem statement**

According to the 2021 Australian Competition and Consumer Commission report, the Culturally and Linguistically Diverse (CALD) community lost $22 million to scams in 2020 [1]. They are more likely to be targeted due to their disadvantage and language skills. On the other hand, it has been reported that in the US, more than 95,000 people reported about $770 million in losses to fraud initiated on social media platforms in 2021 [2]. AI is being used as a tool for scammers, and the CALD community needs more training and assistance than ever. To reduce online harms, the World Economic Forum (WEF) has emphasised "safety by design" principles [3], which involve integrating safety considerations into the design and development of digital platforms. It is essential to consider ways to empower users, promote transparency, facilitate research and access, and set foundational safety standards.

A recent study by [4] shows that people are more likely to be aware of 'post-hoc' safety tools, such as reporting, and content rating, than preventative measures. It demonstrates that there is typically little satisfaction with safety technologies. The usage of such safety tools is more common among those who have witnessed online damage. Concerns regarding these technologies' accessibility for all users are raised because those with higher levels of digital literacy are also more likely to use some of these tools. Therefore, there is a critical need for user-friendly tools that can be used by people who have lower digital literacy, particularly in the context of the CALD community. To address this need, our work aims to design a user-friendly AI-assisted platform called Malak, tailored for users with lower literacy backgrounds to enhance their engagement in online safety tools and provide a comprehensive experience of reducing online harms and healthy online interactions. Our research aims to answer the following questions:

- How can we reduce the harms of AI-generated misinformation and fraud in the CALD population in Australia and, in general, anyone around the world?
- How can we use Generative AI tools to combat disinformation/malicious content to better support the CALD population in Australia and other parts of the world?
- How can soft biometrics with ML mitigate the risks of intimidation, depression, agitation, and psychological stresses due to online interactions?
- How can personalized Generative AI tools assist the CALD population in identifying anomalies in social media, email and other online apps?

## Malak a Multilingual AI assistant

We envision a tailored personal multilingual AI assistant, Malak, to address the above-discussed concerns and help individuals remain safe online. Malak will have five modules to prevent users from engaging in online harm. Figure 1 depicts this assistant.

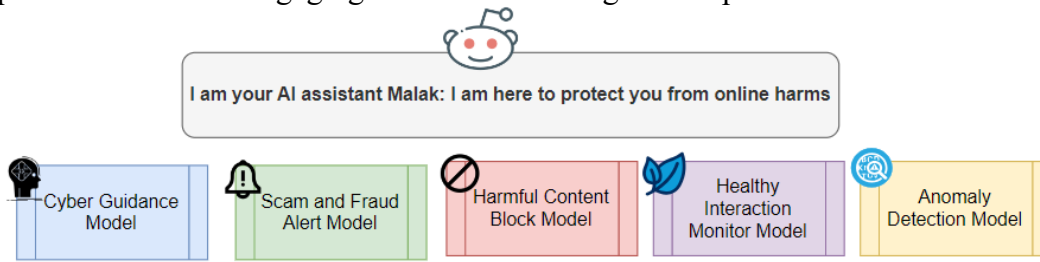

*Figure 1: AI Assistant Malak*

The aspects of each module are as follows:

1. **Cyber guidance model (CGM):** It will guide the users on specific incidents, such as phishing, fake news propagation, etc. The individual selects an option, and then the intelligent agent guides the user through the specific steps. For example, a mature-aged person is heavily involved in social media. Suddenly, the person encountered some fake videos about them on social media. Prompt example of Malak chatting with the individual:

   > Hey Malak, Someone posted a video of me that is not real; what should I do now?

   > Don't worry. I am here to help you. Let me guide you through the process of removing its harms and affects.

   > Based on the pre-trained large language model (LLM), Malak (the Cyber guidance model) will be fine-tuned and adapted to the in-house and publicly available Cyber Awareness and Guidance Dataset.

2. **Scam and Fraud Alert Model (SFAM):** Malak will alert users of all recent frauds on different digital platforms. Malak will provide instructions on how to be safe from these incidents. Example conversation:

   > Hey Zac, have you noticed a new lost family member scam is spreading on platform Z, and many people are falling for it?

   > Oh Malak, I didn't know. Please update me.

   > SFAM of Malak will have subscriptions to Common Vulnerabilities and Exposures (CVEs), reputed Cybersecurity Govt and businesses. The published threat will be automatically analysed by LLM in the digital context of the CALD user, i.e. either email, OS or specific frequent app of the CALD user. If relevant will notify the user.

3. **Harmful Content Block Model (HCBM):** Malak will search for any harmful/violent/extremist content in the individual's feed and the feed of their connection and remove those feeds or tag them as dangerous. Malak analyses the issue based on the provided information. For example, if fake news is in the user's feed, Malak will check and dig out the source, remove that news from the feed, and block the suspect.

4. **Healthy Interaction Monitor Model (HIMM):** Suppose the individual is about to interact with strangers online. In that case, Malak will monitor the users' mental and emotional states using biometric parameters and applying soft biometrics. Suppose it notices unusual biometric parameters such as high blood pressure/unusual heart rate, which might be a sign that the users are engaged in harmful conversations. In that case, Malak will notify the user and ask them to end the conversation. This is an extension of our previous work, where we proposed a usable biometric-based contextual cybersecurity framework to minimize non-intentional behavioural-based malicious cyber incidents measuring people's mental or emotional states using biometric parameters.

5. **Anomaly Detection Model (ADM):** Malak will use anomaly detection techniques to detect suspicious activities on user devices, applications, and social media platforms. Malak will verify unauthorized login incidents, unusual social media, messaging, and app activity in various geographic areas, time of day, duration, routine, and usage profiles to identify anomalies in the digital world of the CALD person.

**Results**: This research will produce a workable prototype with at least two AI models of Malak and test the outcome with the CALD population. The best-case scenario is to design, train and develop all the models and test them with the target groups. The research will use LLM, such as Google Gemini, in addition to Google Colab and Google Cloud, to code and use various AI techniques to implement the functionalities of the Malak AI assistant. In-house and publicly available datasets, such as [5,6,7,8], will be utilized to develop Malak to filter harmful content and guide the users in any reported cyber incident. Malak's AI assistive modules will reduce the rate of phishing attacks and increase cybersecurity awareness. Malak will improve the security of online interaction with vulnerable young children and older adults, especially in the CALD community.

**Literature Review**

Machine learning technologies have been extensively used to classify fake/malicious content, anomaly detection and cyber-attacks. Some recent work [9] has presented the potential of Generative AI tools, such as ChatGPT, for annotating harmful content. However, the work demonstrates that a deeper understanding of the nuances within hateful, offensive, and toxic content is required, and other Generative AI Models need to be tested. Soft biometrics, which analyse an individual's characteristics and emotions rather than identifying them, have been researched for various applications [10]. Researchers have explored soft biometrics for several purposes, including monitoring well-being, assessing psychological profiles and delivering healthcare [11]. Handwriting and signature biometrics have proven effective in predicting emotional states [12]. We extended the usage of soft biometrics in our work [13], where we proposed a usable framework that uses soft biometrics to determine a user's emotional state to minimize non-intentional behaviour-based malicious cyber incidents. One part of this new research aims to extend the framework to bar users from engaging in harmful online discussions. The proposed study also extends our work on anomaly detection using ensemble models [14] and fake news detection using ML [15] and [16]. We intend to publicly share our code and dataset with the broader research community.